# MODIFIED WATERSHED APPROACH FOR SEGMENTATION OF COMPLEX OPTICAL COHERENCE TOMOGRAPHIC IMAGES

Maryam Viqar*, Violeta Madjarova, Elena Stoykova

Institute of Optical Materials and Technologies, Bulgarian Academy of Sciences, 109, Acad. G. Bonchev Str., 1113 Sofia, Bulgaria

**Abstract**

*Watershed segmentation method has been used in various applications. But many a times, due to its over-segmentation attributes, it underperforms in several tasks where noise is a dominant source. In this study, Optical Coherence Tomography images have been acquired, and segmentation has been performed to analyse the different regions of fluid filled sacs in a lemon. A modified watershed algorithm has been proposed which gives promising results for segmentation of internal lemon structures.*

## 1. INTRODUCTION

Recently, tremendous growth is seen in imaging modalities and processing algorithms to make the systems artificially intelligent. In the field of biotechnology and medicine, the commonly used imaging systems are Confocal Microscopy, Computed Tomography Scan (CT-Scan), Magnetic Resonance Imaging (MRI), Optical Coherence Tomography (OCT), Ultrasound (US), etc. Amongst them, the OCT systems have the advantage of depth penetration along with good resolution inside the transparent layers of the sample. Moreover, this is a non-invasive, non-ionising and contactless technique which makes it more desirable. It has been widely used in eye clinics in the past several years for eye related diseases like Macular lesions, Wet Age-related Macular Degeneration (AMD), Papilledema, Retinal Nerve Fiber Layer thickness in Glaucoma patients, Epithelial thickness mapping for detecting corneal abnormalities, Diabetic Retinopathy, Macular Degeneration Retinal layer thickness in early diagnosis of Parkinson's disease and Alzheimer's, etc [1]. Apart from eye clinics, the OCT technique is being explored in other fields like in Microangiography, Cardiology, Otology, Dermatology, Dental Applications, Biotechnological Application like seed differentiation, fruit ageing, etc. Research is done in both domains (hardware and software): (i) to modify the OCT set-up for enhanced acquisition and (ii) for processing algorithms to extract deeper information.

OCT is a high-resolution, contactless, cross-sectional imaging technique capable of acquiring depth information in a transparent object within a range of few millimeters. It is based on the principle of low coherence interferometry. Most commonly used interferometers are fiber-based Michelson type and Mach-Zehnder type interferometers. In Michelson type, the light field is directed to a beam splitter that splits the light into two beams: one illuminates the sample and the other illuminates the reference arm. Then, the reflections from the sample arm and the reference arm propagate back to the beam splitter and get combined to form an interference pattern. The interference will only appear when the two optical path lengths are matched to within the coherence length of the light source. Based on the methods for detection of the interference pattern and post processing to provide information of the in-depth structures having varying refractive indices, the OCT is generally divided into time domain OCT (TD-OCT) and Fourier domain OCT (FD-OCT). In TD OCT each depth scans are acquired by movement of the reference mirror. In FD-OCT, the interference signal is either spectrally resolved at the detector, or a light source that sweeps through a given wavelength range. The obtained spectrally resolved interference signal is then Fourier transformed to generate information of the in-depth structures having varying refractive indices. The three-dimensional scans in OCT are obtained by incorporating multiple depth and lateral X-Y scans. Generally, the resolution of the OCT system is in the range of few micrometers whereas the system can support a depth penetration of a few millimeters [2].

The OCT based images can be used to extract additional information, which is not visible to a naked eye. The images can also be processed to perform segmentation and classification operations for detection/extraction of complex structures which otherwise would consume lot of time and money due to the requirement of trained professionals. To obtain morphological information, several Artificial Intelligence (AI) based methods have been proposed. These models are designed to extract specific layers and structures, to classify diseases in animals, to distinguish age, variety, specie, quality of fruits and vegetables. In this work, volume dataset is acquired for complex internal structures of lemon. The analysis of the internal structure of lemon requires a highly accurate segmentation method which can separate the fluid-filled sacs so that these structures can be used to estimate the age and quality of the lemon fruit in a non-destructive manner. To perform this task, a segmentation method is put forward for segmentation of inner sections of lemon called carpels which are fluid filled (juice) sacs.

Several segmentation techniques have been proposed with application in fields of bio-medical imaging, bio-technology, agriculture, biology, etc. Image segmentation methods usually segregate the regions of interest (objects) in an image based on lines, curves, boundaries, etc. Generally, the techniques are application specific and universal segmentation techniques are yet a challenging task. Commonly used segmentation algorithms are based on following approaches: (i) region expanding methods as in [3]-[11], (ii) thresholding [12]-[13], (iii) graph-based methods [14], etc.

One of the commonly used techniques is Watershed segmentation [3] which is based on topographical features. The image is considered as a surface having varying topographic features relative to the intensities in the image. This surface is filled with water starting with basins created due to higher areas (mountains) surrounding lower ones. The water is made to flow from the lower points in basins (local minima) until it starts merging with the water from different basins, where the watershed lines or dams are built. The water-flow ends when the water reaches to the highest point on the surface.

The watershed transform suffers from over-segmentation problem as huge number of unwanted segments are generated due to noise present and inherent texture details of the image. Hence, several models have been proposed with advancements to avoid the over-segmentation problem. Belaid et. al. in [4] proposed a watershed-based model with adaptive topological gradient method. The method results in attenuation of over-segmentation problem along with less computational complexity making it suitable even for real time applications. A marker-based method presented in [5] uses basic morphological operations by segmenting a skeleton for the edge map. The aim is to keep the fragments of same region together. For the connected segments in the skeleton, the seeds are evaluated by finding out the local minima, to which the dilation process is applied. In this manner, the overlapping regions are combined together and used for segmentation, allowing lesser cancellation or addition of false edges apart from original edge map. Another computationally efficient modified watershed algorithm was introduced by Chien et. al. in [6] for segmentation in videos. This method exploits the temporal coherence and only modifies watersheds for each next frame rather than searching for watershed in all the frames. Furthermore, it uses Intra-inter watershed scheme wherein the I-watershed frames are inserted to maintain accuracy and avoid error propagation. An alternative to watershed for brain tumours is proposed by Aslam et. al. [7] which follows the motion of a falling ball on the surface having varying topography. The method is a region growing segmentation technique which uses Fuzzy logic for decision making.

These segmentation methods suffer from several problems, such as application specific nature, computational complexity, over or under segmentation, etc. Hence, depending on the image, the amount of noise present, the textural details, the above-mentioned methods fail to give optimum performance. Typical structures like that of lemon fluid sacs are complex in nature and moreover noise is commonly experienced phenomenon in the images acquired using the OCT set-up due to back-scattering of light. Hence, in this work a method is proposed to overcome the problem of unwanted segmentation due to noisy and inherent textural details of an image employing the watershed transform as the back-bone. The novelty of the work lies in its segmentation accuracy even in the presence of noise and the ability to limit over-segmentation.

The proposed model firstly pre-processes the image in three stages: Stage I where Hann transform is applied on raw A-scan to smoothen the effect of edges and corners; Stage II wherein pre-processing using the contours followed by gradient and marker-based method is used to segregate the foreground and background using the morphological operations; in Stage III the watershed transform is applied to segment into regions based on topography. The pre-processing stages involving the Hann operator and marker-based technique respectively, enhance the performance of the conventional watershed method. It identifies the real edges efficiently and henceforth, helps to avoid over-segmentation.

**Fig. 1** (a) Schematics of the acquisition set-up for image acquisition of lemon (b) Structure of lemon showing the external and interior portions in lemon. Internal structure consisting of fluid filled sacs

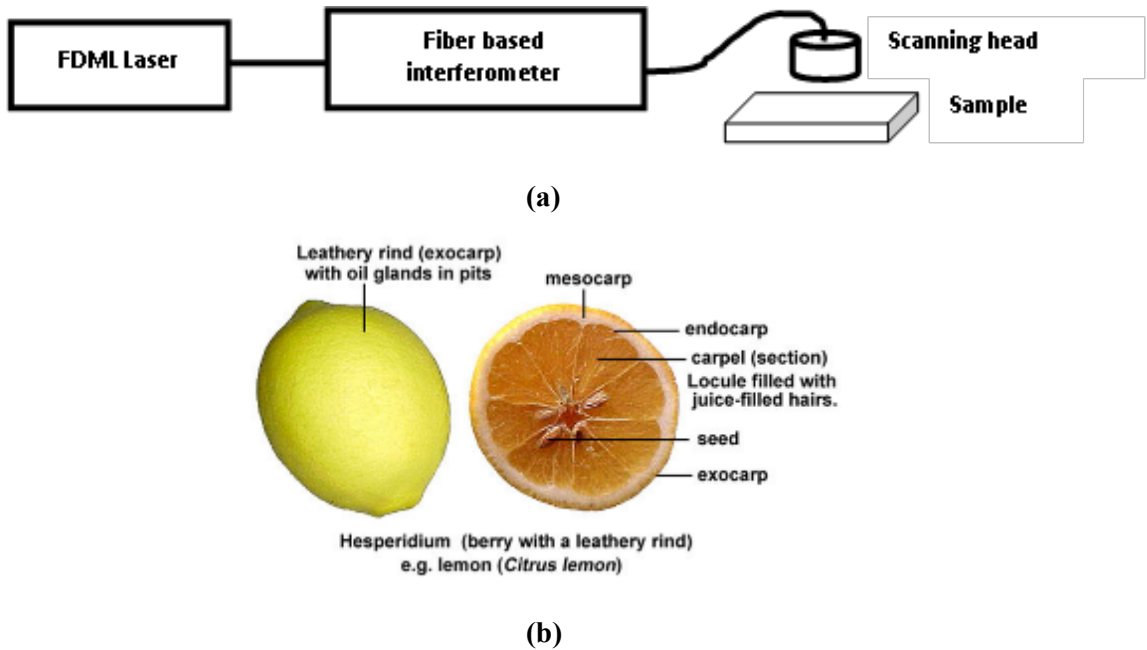

(a)

(b)

The remaining section of the paper contains the following information: Section II describes the watershed transform and the proposed modified watershed transform model, Section III gives the results Section IV provides discussion and comparison of the results, Section V gives concluding remarks., followed by list of references.

## 2. MATERIALS AND METHODS

*2.1. OCT Set-up and lemon*

The used OCT setup is produced by Optores in Munich, Germany [15]. It is a FD OCT system with Fourier Domain Mode Lock (FDML) laser [16] which operates at 1.6 MHz sweeping rate with 100 nm sweeping range at central wavelength 1309 nm. The depth resolution is calculated to 17 micro-meters and a lateral resolution to 40 micro-meters. The schematics of the acquisition set-up is given in Fig. 1 (a). The area scanned is of size 7 mm x 7 mm (1024 by 1024 points). The sample arm of the fiber-based interferometer is connected to a scanning head via an optical fiber. The scanning head performs the X-Y scanning. In this FD-OCT, the source sweeps through an optical bandwidth, while the detector acquires signal at each evenly distributed wavelength. This spectrally resolved signal is Fourier transformed to get depth-resolved profile. The line scan acquisition is done at several points by laterally scanning on the sample in X direction to get the complete 2D depth-image (B-scan). The volume scan is acquired by performing additional scan in Y direction.

A fresh lemon was brought from Local Vegetable and Fruits market in Sofia, Bulgaria. It was peeled, placed on a Petri-dish on a height-adjustable table. The volume dataset was acquired using the OCT setup as shown in Fig.1 (a). while the lemon structure can be seen in Fig. 1 (b) and can be described as follows: (i) the outer layers of the lemon are exocarp and mesocarp which are followed by an inner part called the endocarp; (ii) the inner endocarp is segmented into main lobes by carpel which contains the juice filled fluid sacs. The fluid filled sacs contain the juice which is the most important part of the lemon as afar as edibility of the lemon is concerned. These sacs are of irregular shape and size making an overall complex structure to segment.

*2.2. Methodology*

**Fig. 2** Block diagram of the proposed modified Hann windowing and marker-based watershed segmentation

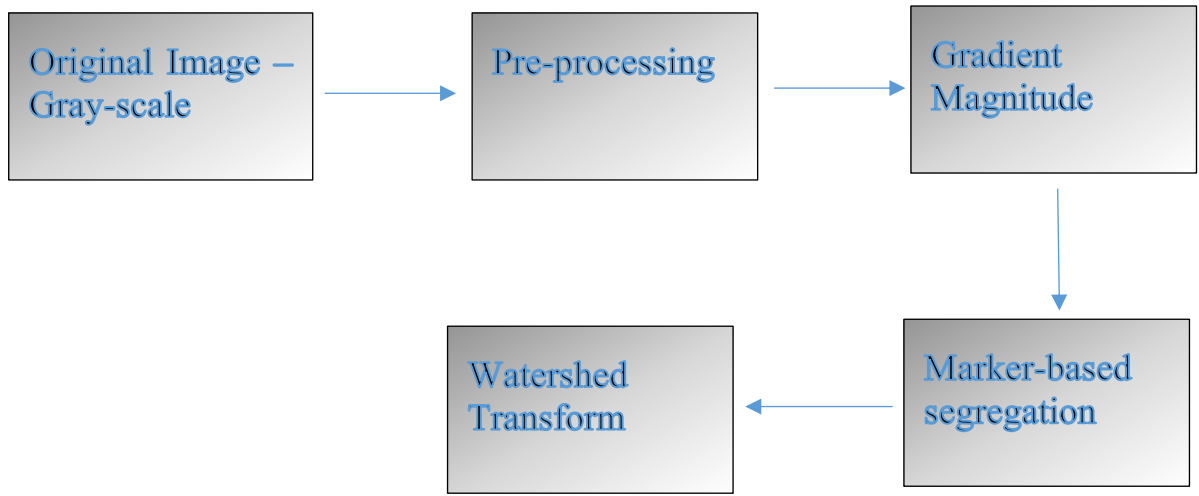

The used watershed algorithm is based on the topological representation of an image. The image is assumed as surface flooded with water, wherein the water flows towards the lower areas as a result of gravitational force. Three regions are identified this way: (i) Minima, (ii) Watersheds and (iii) Catchment basins. The altitudes of the topological surface are designated using the gradient of the image. When water flows, firstly it goes towards the lowest points of the catchment basin. Furthermore, as the water is allowed to flow, the water reaches the top of catchment basins and may start to merge with neighbouring flows. This identifies the watershed, which corresponds to details of the image having high gradient information, which can be employed to segregate the regions of an image. The process of water-flow is immersion type in which the water emerges out of the spots at the minima of the basins and it over-flows the height of surrounding boundaries. The points of merger of this over-flow water are the lines where the dam is to be built. Upon flooding the complete surface with water, these dams are the watershed lines of the image.

There are many implementations of the watershed algorithm [3]- [8] amongst which Vincent and Soille's algorithm [3] is computationally less complex and is described as below:

- Firstly, the gradient of the complete image is calculated.

- These gradient values are used to sort the pixels values of the image, starting with the lower gradient values. Special tags are given to pixels of connected areas.

- First the lower gradient values are processed following the higher ones. On each of the gradient levels, pixels with tagged neighbours are inserted into a priority pipeline first. Initial tag of the pixel is decided by its neighbours in the pipeline. After processing of the tagged neighbouring pixels, untagged ones are also tagged.

- After the tagging of all the pixels, the outlines separating the regions with different tags are the watershed lines of the image.

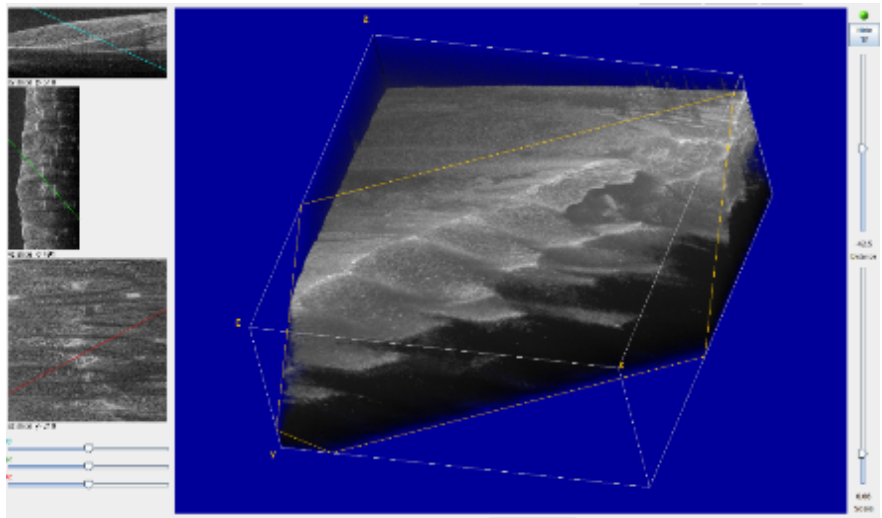

Fig. 3 3D OCT image of lemon, sliced at plain marked with yellow contours

Mathematically, the watershed can expressed as follows: Let the image be w(x,y), with (x,y) ∈ $\mathbb{R}^2$, $\Omega$ is the open-bounded domain of $\mathbb{R}^2$. The pixels c of $\Omega$ represent the pixels in catchment basin D(M) having M as the minima. The threshold set of w for a height R be $T_R = \{c \in \Omega, w(c) \leq R\}$, and $R_{min}$ and $R_{max}$ are the minimum and maximum values of image w. The level of R increases from its minimum to maximum value, which results in expansion of basins. The union of sets of basins calculated at step R, are considered as $Y_R$. The union of all the minima corresponding to different regions for height R, the watershed can be defined using the following recursion:

$$\{Y_{R_{min}} = T_{R_{min}}, \forall R \in [\ R_{min}, R_{max} - 1], Y_{R+1} = min_{R+1} \cup IZ_{T_{R+1}}(Y_R)\} \qquad (1)$$

In eq.1, $IZ_{T_{R+1}} = \bigcup_{i=1}^{n} iz_{T_{R+1}}(Y_{R_i})$, where n is number of minima of gray-level image I and the $iz_{T_{R+1}}(Z_i)$ is defined in eq.2 as follows:

$$iz_{\Omega}(Z_i) = \{z \epsilon \Omega, \forall n \neq i, d_{\Omega}(z, Z_i) \leq d_{\Omega}(z, Z_n)\} \qquad (2)$$

For a grey-level image I, set $Y_{R_{max}}$ is equivalent to set of catchment basins. The compliment of the $Y_{R_{max}}$ in $\Omega$ gives the watershed of the image [4]. The watershed algorithm offers many benefits, firstly it results in to closed segments which are important in segmentation tasks and secondly it requires lesser time for processing compared to many of the state-of-art methods. But in addition to these advantages, there are also disadvantages associated with this method, over-segmentation being the most problematic one. The marker-based segmentation technique is an example of over-segmentation avoiding technique. In this work, we propose a pre-processing along with employing the marker model for the watershed transform to solve the problem more effectively.

The pre-processing model makes use of Hann windowing which is a window function used for tapering the high-frequency features in an image. It is applied after the acquisition of A-scan prior to Fourier Transform. The OCT scans are pre-processed using the thresholding with a value of 245 (chosen heuristically), filling of holes and use of Chan-Vese and Edge based active contouring to get a binary image. The marker-based foreground and background demarcation followed by the watershed algorithm is used to segment the internal structures of lemon. The block diagram of the proposed scheme is presented in Fig. 2.

The marker model works by identifying the foreground and background objects. The foreground objects are marked as connected sets of pixels covering the region of an individual object. One of such techniques is to use morphological operations namely reconstruction-based opening and closing. While using this operation, the markers reach even up to the level of the edges of the objects. Hence to erase the edges from the marked fore-ground regions, closing operation is followed by an erosion process.

Furthermore, to mark the background regions basic thresholding process is done. But this process, marks down the whole background region whereas the desired result includes only the markers for the background region. To get these markers, the "skeleton by influence zones" technique is used. Lastly, minima are computed for regions of objects marked as foreground and background. This sums up the process of the marker, which is followed by the watershed segmentation described above [5].

**Fig. 4** (a) Original OCT image of lemon (b) Pre-processed binary image (c) Gradient Magnitude
(d) Watershed lines (e) Original watershed segmentation image (f) Proposed watershed

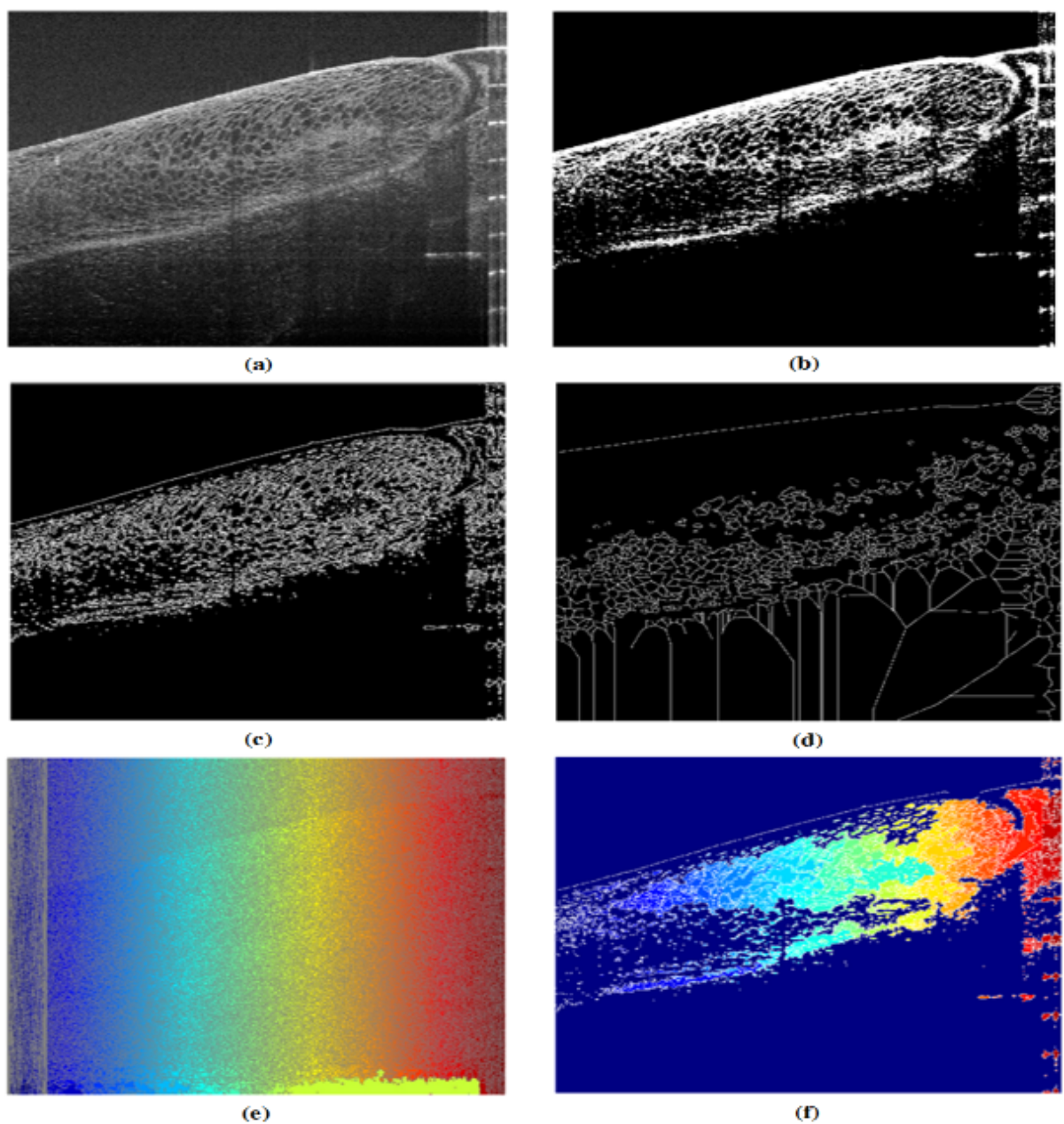

The watershed along with the marker-based foreground and background pre-processing is known as the marker-based watershed transform. Though it aims to avoid the over-segmentation problem it does not address the problem efficiently. Hence, in this work more pre-processing techniques are used along the marker-based segmentation to attain the desired result for the segmentation of complex structures as seen in a lemon.

The over-segmentation in the watershed algorithm is present many times due to edge-based artefacts which are very common in OCT images. In such cases, to supress these edge effects the use of the Hann window function prior to segmentation followed by pre-processing using markers and contours is proposed in this work. The Hann window function is one such window which reduces the impact of the edge and corner by tapering the image-based signal and can be expressed as follows:

$$W_{Hann}(i,j) = \frac{1}{4}(1 - \cos(\frac{2\pi i}{P-1}))(1 - \cos\left(\frac{2\pi j}{Q-1}\right)) \qquad (3)$$

In eq.3, i, j, P and Q represent the current horizontal and vertical positions, width and height of a block respectively. This function is a 2-dimensional, which increases the efficiency of the segmentation method by prioritizing the central regions of an object (to be segmented) rather than their edges and corners [15].

## 3. RESULTS

The results of the proposed and original watershed model along with inter-mediate steps are presented in Fig. 4. The original 3D OCT volume data are given in Fig. 3 and one slice of the original image having size 1024 × 900 pixels (7 mm by 5 mm) is shown in Fig. 4a as captured by the OCT system. The volume dataset consists of several B-scans, where each B-scan was extracted as an individual 2-D image and modified watershed processing was performed on these scans. Each single image is a grey-scale image afflicted with noise. The binary image after the pre-processing contour-based operation is represented in Fig. 4b, where we can observe a binary image with showing different regions. Furthermore, Fig. 4c shows gradient estimation prior to marking of foreground and background regions. The watershed lines which separate regions based on their topographic characteristics are shown in Fig. 4d. The regions of fluid filled sacs detected by the modified watershed algorithm are shown in Fig.4f. It can be visualised that the modified approach can efficiently segment the sacs filled with lemon fluid demonstrating clearly the edges of each sac. In contrast to this, the results of original watershed segmentation in Fig. 4e exhibit over-segmentation. In Figs. 4e and 4f, the colours are for better visualization of the labelled regions; the original watershed's outcomes incorporate the noise into false edge and produce regions for original edges, noise generated edge and combination of both.

In summary, the segmented area are blobs of pixels marked with outliners (marked with white) to separate foreground from background region. Initial pre-processed image works as a foreground marker but it cannot separate out the fluid containing areas from the boundaries of the sacs. For extracting the edge-map of the fluid filled regions, further marker-based watershed transform is applied. It is able to extract the edges of the fluid filled sacs leaving behind the fibrous structures and skin area of lemon which do not have useful fluid content.

Generally, most of image acquisition techniques like OCT in this work are afflicted with different kinds of distortions due to surrounding environment, capture mechanism, transmissions, mathematical processing algorithms. It is highly difficult to obtain distortion-less images. The time required for watershed transform is 0.18 seconds, whereas the time for the proposed modified watershed is 2.13 seconds. The time complexity has increased due to additional pre-processing algorithms but the segmentation outputs are far more practically applicable. Hence, methods must be such that even in presence of noise, they should be capable enough to segment accurately.

## 4. DISCUSSION

The modified approach which in addition to watershed segmentation, includes gradient, contour-marker and Hann windowing approaches is quite capable to segment the fluid filled sacs even in the presence of false edges and corners generated due to noise.

The watershed approach is good when the image has only true edge and no effects due to external sources. The Hann-window works well to taper the effect of edges, and results into an image with dominant edges, as usually the edges due to noise do not have pronounced effect. The marker-based method further helps in marking the regions of interest using the foreground and background technique effectively. Lastly, the watershed technique when applied to such pre-processed images, results into appropriate segmented areas.

The accurate segmentation results come at the cost of higher computational complexity as the time for processing increased by several folds.

## 5. CONCLUSIONS

The watershed technique is powerful for segmentation, but has certain limitations due to presence of artificially created points, corners, boundaries, edges, etc. The technique is not well-suited for images afflicted to distortion of types speckle noise, blocking artifacts, Gaussian noise, etc. The use of Hann windowing helps to smoothen and supress the effect of edges. This pre-processing step helped significantly to segment the lemon sacs more accurately avoiding the presence of undesired segmented areas arising due to distortions. These methods can in future be used to analyse the citrus fruits like lemon to estimate the quality, life or juice content. More processing would be required to relate the properties with the segmented regions as many times large segmented fluid sacs will not depict proportionality to juice content. The skin structures can be examined and automated methods can be developed for biotechnology to estimate the ageing in the lemon fruit.

## ACKNOWLEDGMENTS

V.M. European Regional Development Fund within the Operational Programme "Science and Education for Smart Growth 2014–2020" under the Project CoE "National center of Mechatronics and Clean Technologies" BG05M2OP001-1.001-0008. M.V. would like to thank European Union's Horizon 2020 research and innovation programme under the Marie Skłodowska-Curie grant agreement No 956770 for the funding.